\documentclass[journal,onecolumn]{IEEEtran}

\usepackage{graphicx}
\usepackage{wrapfig}
\usepackage{romannum}
\usepackage{empheq}
\usepackage{amsmath}
\usepackage{amsfonts}
\usepackage{amssymb}
\usepackage{amsthm}
\usepackage{lipsum}
\usepackage{caption}
\usepackage{bm}
\usepackage{mathtools}
\usepackage[noadjust]{cite}
\usepackage{float}
\usepackage{xcolor}

\usepackage[ruled,vlined]{algorithm2e}
\usepackage[noend]{algpseudocode}
\usepackage{romannum}
\usepackage{color}
\newcommand{\etal}{\textit{et al}.}
\title{D3PG: Dirichlet DDPG for Task Partitioning and Offloading with Constrained Hybrid Action Space in Mobile Edge Computing
}

\author {Laha Ale, Scott A. King, Ning Zhang, Abdul Rahman Sattar, Janahan Skandaraniyam

}
\begin{document}

\maketitle

\begin{abstract} 
 Mobile Edge Computing (MEC) {\color{black}has been regarded as a promising paradigm to reduce service latency for data processing in Internet of Things, by provisioning computing resources at network edge. In this work, we jointly optimize the task partitioning and computational power allocation for computation offloading in a dynamic environment with multiple IoT devices and multiple edge servers. We formulate the problem as a Markov decision process with constrained hybrid action space, which cannot be well handled by existing deep reinforcement learning (DRL) algorithms. Therefore, we develop a novel Deep Reinforcement Learning called Dirichlet Deep Deterministic Policy Gradient (D3PG), which 
is built on Deep Deterministic Policy Gradient (DDPG) to solve the problem. The developed model can learn to solve multi-objective optimization, including maximizing the number of tasks processed before expiration and minimizing the energy cost and service latency.} More importantly, D3PG can effectively deal with constrained distribution-continuous hybrid action space, where the distribution variables are for the task partitioning and offloading, while the continuous variables are for computational frequency control. Moreover, the D3PG can address many similar issues in MEC and general reinforcement learning problems. Extensive simulation results show that the proposed D3PG outperforms the state-of-art methods.

\end{abstract}

%Firstly, to feed the deep learning model with more data and improve the robustness of the model, data augmentation is performed.

\IEEEpeerreviewmaketitle

\section{Introduction}
\label{sec_intro}

Internet of Things (IoTs)~\cite{IoT9260173,zhang2020physical} is considered as the foundation for a wide range of applications, including self-driving cars, smart cities, and environment monitoring. Although IoTs can address small tasks with a reasonable amount of energy consumption, many computational-intensive tasks are beyond the capacity of the IoTs. Moreover, many applications, such as self-driving cars and smart factory robots, require real-time responses, and the IoTs struggle to respond to users when they require a relatively large amount of computational resources. Furthermore, most of the IoTs are extremely sensitive to energy consumption when these devices are running wireless. Since IoTs have limited computational resources and energy support, they offload computational- and delay-intensive tasks to online servers to process the tasks. However, it is challenging to offload a large number of tasks through the core network and process them on remote servers because the networks would be congested and therefore increase the delay time. Mobile Edge Computing (MEC) is proposed to address the tasks in proximity and reduce the burden of the core network~\cite{chen2019energy}. Unfortunately, MEC is not a panacea for the above problem because MEC servers are equipped with much less computational resources than central cloud servers; therefore, task offloading and scheduling optimization are vital to exploit the limited resources and improve service quality and reduce costs. 

{\color{black}
Various methods have been proposed to optimize the MEC resource usage to fully utilize the limited computational resource in MEC servers. To reduce the idle time of edge servers and respond to IoT user timely, an offloading task can be sliced into small sub-tasks, and the sub-tasks can be processed in heterogeneous edge servers~\cite{li9119487}. Therefore, slicing task~\cite{9024603} into small sub-tasks and offloading to edge servers so that the limited computational power on the edge servers can be fully utilized. Conventional optimization methods (eg. CVX and MIP)~\cite{9099242}, machine learning~\cite{8292514}, deep learning~\cite{Lecun2015}, and reinforcement learning~\cite{SuttonRL,dqn_nature} methods have been introduced to address computation offloading challenges. It is very challenging to adopt the aforementioned methods into MEC tasks partitioning and scheduling. First, it is considerably challenging to describe the practical MEC network into mathematical forms that can fit into conventional optimization methods. Second, machine learning and deep learning usually require labeled data to train the models, which can be  extremely difficult for humans to manually compute and label slicing and scheduling datasets. 

Deep reinforcement learning (DRL) methods can mitigate the issues mentioned above. To optimize the task partitioning, offloading, and computing power allocation, the existing methods typically use DRL and optimization techniques to deal with those decision variables separately, which can lead to a poor overall system performance, instead of addressing the joint optimization problem in an end-to-end manner. To jointly optimize those decision variables, we need to deal with a hybrid action space. Moreover, it is even more complex as there are some constraints on the action space, because the sum of the percentage of all sub-tasks for offloading from a given task should be equal to one. The majority of existing DRL models can only address discrete action~\cite{dqn_nature} or continuous action space~\cite{Lillicrap2016}. Several works try to address hybrid action with approximation or relaxation of continuous action space~\cite{ale2021}. Hausknecht~\etal ~\cite{Hausknecht2016}. However, they cannot address constrained action space in edge computing. Wu~\etal~\cite{Wu9277604} use $softmax$ to capture the task partitioning actions to satisfy the constraints of the action space, which is a proportional action space, the sum of which should be one. However, $softmax$ does not have an exploration mechanism to explore all the possible actions and derive the optimal policies.

%Moreover, the existing methods require further optimization based on the outputs of the learning models instead of addressing the joint optimization problems in an end-to-end manner. Finally, it is challenging to extend the current reinforcement learning or DRL methods to address fine-grained task partitioning and offloading simultaneously in MEC. It is even more complex as the action space is coupled with constraints.  Wu~\etal~\cite{Wu9277604} use $softmax$ to capture the task partitioning actions to satisfy the constraints of the action space, which is a proportional action space, the sum of which should be one. However, $softmax$ does not have an exploration mechanism to explore all the possible actions and derive the optimal policies.

% However, it is challenging to divide the tasks into sub-tasks and offload them to the edge servers based on the configuration of the edge servers and network conditions~\cite{9024603}.

In this work, we propose  a  novel  deep  reinforcement  learning  approach called  Dirichlet Deep Deterministic  Policy  Gradient  (D3PG), based on Deep  Deterministic  Policy Gradient (DDPG), to jointly optimize the task partitioning, computation offloading, and computational frequency control. The developed model can decide to partition the tasks flexibly, offload the sub-tasks to the edge servers, and  select the computational frequencies of edge servers to execute the sub-tasks. The goal of the model is to make those complex decisions based on the observation to maximize the number of tasks completed before expiring and minimize the energy consumption and time cost. The model can make the decisions and jointly optimize multiple objectives. The main contributions of this work are:

\begin{itemize}
 \item We developed a novel D3PG model to optimize MEC resource allocation and improve service quality. The proposed model generates a distribution-continuous hybrid action space to address various issues flexibly. Specifically, each action includes a distribution formulated as a Dirichlet distribution for partitioning and offloading tasks, and continuous components for frequency control.
 
%  \item We use Mean Squared Error (MSE) of Q-values and Kullback–Leibler (KL) divergence to design a new loss function that boost the learning process. We believe this model can be generalize to address many other reinforcement learning problems. %Additionally, prioritized experience replay has been introduced boost learning and fully use the rare but important samples to avoid the model over-fitting on the common samples.
 
 \item A configurable optimization target is proposed to address multiple joint optimization problems. The model optimizes multiple objectives in an end-to-end manner, and it does not require further optimization like existing methods. 
 
 \item We have tested the developed method by extensive simulations, and results show our method outperforms the the state-of-art methods.
 
\end{itemize}
}
The rest of the paper is organized as follows. Section~\ref{related_work} investigates related works. Section~\ref{sec_sys_model} presents the system modeling and problem formulation. Section~\ref{sec_prop_method} introduces the proposed method in details. Section~\ref{sec_sim_res} provides the simulation  results and Section~\ref{sec_con} concludes this work.

\section{Related Works}
\label{related_work}

In the literature, many methods have been proposed for task scheduling and offloading.  A joint computation offloading and system resource allocation for MEC has been formulated as a mixed-integer non-linear programming format in~\cite{bi_mip}. Then the authors transformed the non-linear programming to linear programming to reduce the complexity and tackle the challenges. Liang \etal~\cite{Liang2019}, adopt linear-fractional programming (LFP), a generalization of linear programming (LP), and a greedy algorithm to optimize the offloading rate and energy consumption. The tasks can also be divided into sub-task processes and processed on local devices or edge servers simultaneously~\cite{9120484,8616760} based on optimization. Similarly, Gao \etal~\cite{Cao8462758} propose a method to find the optimal ratio of task partition into two sub-tasks for edge servers and local IoTs. He \etal~\cite{He9039590} propose an optimization method to partition deep learning inference tasks and offload the partitions to edge servers to find optimal-delay for computing resource allocation. Those standard optimization methods are relatively straightforward to develop and have addressed many optimization problems in task offloading. However, MEC network environments are far complex to describe with mathematical forms, and it is considerably challenging to extend conventional optimization to high-dimensional observations. 

Machine learning and deep learning~\cite{Lecun2015} models can learn from historical labeled data and predict future computational offloading so that the system can make plausible decisions for MEC. Shen \etal~\cite{Shen8954683} surveyed machine learning methods for resource slice and planning for the next-generation network. They have also summarized various machine learning and deep learning methods adopted in computational offloading and content offloading for MEC. Lyu \etal~\cite{Lyu2019} use stochastic gradient descent, a popular machine learning training method, to learn and partition data to offload spatially distributed edge servers; the authors argue that the proposed method can make optimal decisions for data partitioning with respect to time delay. Yang \etal~\cite{Yang9031741} proposed a statistical machine learning method to minimize the energy consumption for edge inference~\cite{Xu2018,Xu8632695}, where deep learning inference tasks are processed on MEC. Ale \etal~\cite{ale8660445} introduced a deep recurrent neural network to capture and predict the user requests so that they can make decisions for the content offloading and allocate resources based on the prediction. Summarily, the computation offloading problems can be formulated as a supervised classification problem and minimize the cost using deep learning~\cite{8292514}. The deep learning models can also adopt bandwidth allocation optimization and maximize the system utility~\cite{9099242}. Machine learning and deep learning methods can learn and predict the offloading decisions optimized with respect to exploiting resources and reducing the cost for the MEC. However, machine learning and deep learning require training data with labels, which require enormous effort to collect and label data. Moreover, it is challenging for humans to label data from optimization problems because it is difficult to take optimal actions in such a complex system.
 
To mitigate the above issues, reinforcement learning~\cite{SuttonRL} and Deep Reinforcement Learning~\cite{dqn_nature,lale9380662} methods are extensively adopted for resource planning and optimization in MEC. In the reinforcement learning framework, we do not need to provide labeled data to train the models; instead, the learning agents interact with the environment (i.e., the MEC networks) to learn and find the optimal policies with respect to the objective function (e.g., minimize energy consumption). A Q-learning (a typical reinforcement learning algorithm) based method has been proposed~\cite{9024603} to make decisions for task offloading; the agents learn to decide whether the current tasks are offloaded to edge servers or processed on local devices to minimize the delay time. Similarly, deep Q-learning has been adopted to decide task offloading and select targeted edge servers for smart vehicles~\cite{8660505}. Cheng \etal~\cite{Cheng8672604} adopted the DRL method to minimize the time delay for computation offloading to Unmanned Aerial Vehicle (UAV) based edge servers and the DRL model outperforms brute-force and the greedy algorithm. Similarly, Baek \etal~\cite{Baek9141290} proposed a Deep Q-Network (DQN)~\cite{dqn_nature} variant by replacing the convolutional neural network with a recurrent neural network to control and select task offloading to edge servers; the offloading actions generated by policies to exploit the limited MEC resources and maximize number of tasks been processed before expiring. Another DQN variant model~\cite{9139263} is proposed to resource allocation optimization by incorporating Bayesian learning and Long Short-Term Memory (LSTM)~\cite{HochSchm97}. In~\cite{li9119487}, the authors have adopted DRL models to optimize task partitioning and scheduling for vehicular networks that allow two edge servers to process a task collaboratively. Yu~\etal~\cite{Yu9205252} proposed a DRL model to optimize the task partitioning and offloading; the model makes decisions based on the profiles of sub-tasks and chooses the local devices or edge servers to process the sequentially depended sub-tasks. 

Although reinforcement learning and DRL methods are adopted to address many task offloading and resource allocation problems in MEC, the currently existing methods can only deal with relatively small action spaces and are inflexible to partition large tasks. For example, the most of the DRL models for MEC can only take binary action that decides a task to offload to the MEC server or process on local devices. Another type of DRL model can find an optimal proportional (percentage) task to offload to edge servers to process. {\color{black}In addition, the majority of reinforcement learning models deal with discrete action~\cite{dqn_nature} or continuous action space~\cite{Lillicrap2016}. Several works try to handle hybrid action with approximation or relaxation of continuous action space~\cite{ale2021}. Hausknecht~\etal ~\cite{Hausknecht2016} relaxed the action space to have to support the hybrid action. Masson~\etal~\cite{Masson2016} handled discrete action with Q-learning and policy search for continuous action. Similarly, Khamassi~\etal~\cite{Khamassi2017} use Q-learning and policy gradient to achieve the same results. Those methods assume on-policy and handle discrete and continuous actions separately. Xiong~\etal~\cite{Xiong2018} and Fu~\etal~\cite{Fu2019} use a hierarchical structure~\cite{Raza2019} to deal with the discrete actions and generate the continuous action based on the discrete actions. Neunert~\etal~\cite{Neunert2020} provided mixed policy to handle the discrete-continuous action space. However, none of those methods can handle constrained action space well in edge computing. 

}

\section{System Model and Problem formulation}
\label{sec_sys_model}

\subsection{System Model}

\begin{figure}
  \centering
  \includegraphics[width=5.5in]{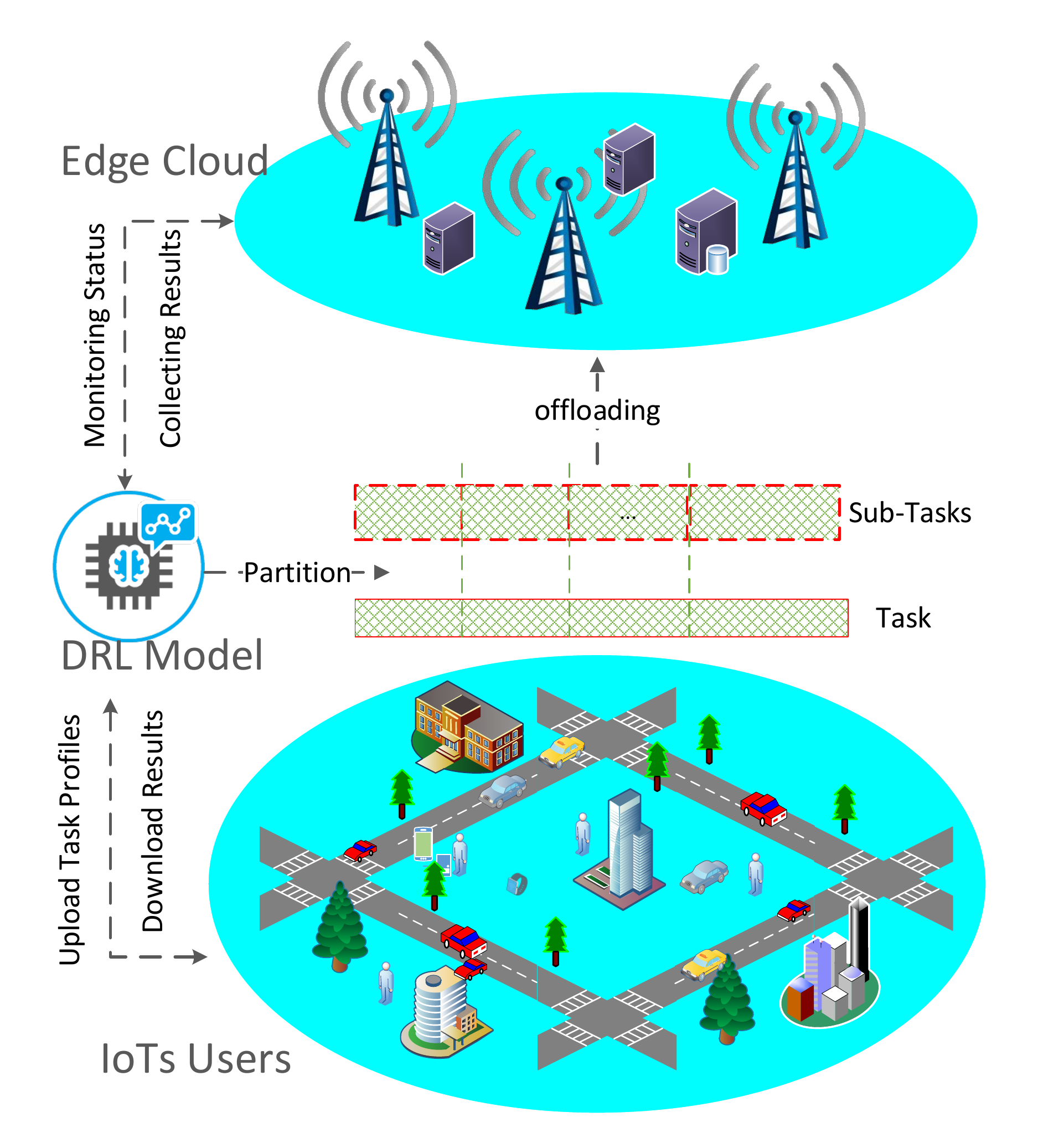}
  \caption{System Model}
  \label{fig:sys_model}
\end{figure}
As shown in Fig.~\ref{fig:sys_model}, there is a group of $N$ IoT users $\mathcal{U} = \{u_1,\dots,u_N\}$ during a given time slot $t$. {\color{black}The IoT devices rely on a set of $K$ MEC servers $\mathcal{M} = \{m_1,\dots,m_K\}$ to process tasks through computation offloading.} A task from the $i^{th}$ user can be defined as: $\Omega_i=\{\mathcal{D}_i,\mathcal{C}_i,\Delta_{max}\}$, where $\mathcal{D}_i$, $\mathcal{C}_i$, and $\Delta_{max}$ denote the data size of the task, required CPU cycles to compute the task, and maximum tolerant latency (expired time) of the task $\Omega_i$, respectively. Further, the tasks can be partitioned into smaller sub-tasks and offloaded to different MEC servers for parallel processing. Specifically, {\color{black}a task $\Omega_i$ can be partitioned into several smaller sub-tasks, $\boldsymbol{\xi}_i = (\xi_{i,1},\dots,\xi_{i,j},\dots, \xi_{i,K})$, where $j$ is the index of the sub-task, and the number of sub-tasks is no greater than the number of MEC servers $K$. It takes $\delta_{i,j}$  to complete each sub-task $\xi_{i,j}$. In other words, the time cost of all the sub-tasks of $\Omega_i$ can be denoted as a vector $\boldsymbol{\delta}_i = (\delta_{i,1},\dots,\xi_{i,j},\dots, \delta_{i,K})$, $i \le N$, and $j \le K$, and the time cost $\delta_{i,j}$ for sub-task $\xi_{i,j}$ can be computed by}:
\begin{equation}
\label{eqn_waiting}
\delta_{i,j} = \delta_{i,j}^{T}+\delta_j^R + \delta_j^Q +\delta_{i,j}^{C},
\end{equation}
{\color{black}where $\delta_{i,j}^{T}$,  $\delta_j^Q$, $\delta_{i,j}^{C}$, and $\delta_j^R$, are the transmission time, queuing time, the computing time of the sub-task $\xi_{i,j}$, and the remaining running time for the task being processed at the server, respectively.}

The transmission time $\delta_{i,j}^T$ can be given as below
\begin{equation}
\label{eqn_tranfer_time}
\delta_{i,j}^T = \frac{D_{i,j}}{\zeta_{i,j}},
\end{equation}
where $\mathcal{D}_{i,j}$ is the data size of the $j^{th}$ sub-task, and $\zeta_{i,j}$ is the current transmission rate from the $i^{th}$ user to the $j^{th}$ MEC server. The transmission rate $\zeta_{i,j}$ can be 
given by
\begin{equation}
\label{eqn_speed}
\zeta_{i,j} = \mathcal{B}_j log_2(1+\frac{P_{i,j}h_{i,j}L_{i,j}}{N_0}),
\end{equation}
where $\mathcal{B}_j$ is the bandwidth, $P_{i,j}$ is the transmission power, and $h_{i,j}$, $L_{i,j}$, $N_0$ are the Rayleigh fading, path loss, and noise power, respectively.

{\color{black}Once the sub-task $\xi_{i,j}$ arrives at the edge server after transmission, it needs to wait for processing before all the sub-tasks in the queue and the server are completed.} $\delta_j^R$ is the reminding computing time of the sub-task which is being processed by the $j^{th}$ MEC server, and it is {\color{black} the difference between the estimated completion time $\delta^C$ and the starting  time $\delta_{s}$} :
\begin{equation}
\label{eqn_waiting_run_time}
\delta_j^R = \delta^C - \delta_{s}.
\end{equation}
{\color{black}
where $\delta^C = \frac{\mathcal{C}_{k}}{f_{k}}$ , and $\mathcal{C}_{k}$ is the required CPU cycles to compute the $k^{th}$ offloaded sub-task in the current edge server and $f_k$ is the allocated computing frequency to process this sub-task.
Further, assume there are $J$  tasks or sub-tasks offloaded to the $j^{th}$ edge server before the current sub-task is assigned to the $j^{th}$ edge server. Then, the waiting time in the queue $\delta_j^Q$ for the $j^{th}$ sub-task on the $j^{th}$ MEC server can be computed by:
\begin{equation}
\label{eqn_queue_time}
\delta_j^Q = \sum_{k=1}^{J}\delta^C_j,\ 
\end{equation}
}

{\color{black}
After all the sub-tasks before the current task are completed, the sub-task $\delta_j^C$ can be processed now. The required computation time $\delta_j^C$ for the $j^{th}$ sub-task on the $j^{th}$ MEC server is given by:
\begin{equation}
\label{eqn_queue_time2}
\delta_j^C = \frac{\mathcal{C}_{i,j}}{f_{i,j}},\ \mathcal{C}_{i,j} \in \mathcal{C}_i,
\end{equation}
}where $\mathcal{C}_{i,j}$ is the required CPU cycles to compute the $j^{th}$ sub-task of $\Omega_i$, and $f_j$ is the frequency of $j^{th}$ MEC server for processing the sub-task.

A task $\Omega_i$ is considered to be completed before the corresponding deadline if all sub-tasks are completed no later than the maximum tolerant latency. In other words, if the last sub-task has been processed before the expiration time, then the task is processed successfully; otherwise, it is considered as expired and a failure to respond to the user. {\color{black}We can denote a positive flag $(+1)$ to the task partitioning and offloading coordinator when the task is completed before the corresponding deadline, and set this flag $0$  when it fails to timely respond to the user, as defined:}
\begin{equation}
\label{eqn_time}
\Lambda =
\left\{
\begin{array}{ll}
  { +1}, & {\text { if } max(\boldsymbol{\delta}_i) \leq \Delta_{\text {max }}}\\
  0, & {\text{ otherwise}}.
\end{array}
\right.
\end{equation}
Note that the agent gets a punishment (energy cost) whenever it takes action for the partitioning and offloading; therefore, the agent gets negative feedback when it fails to respond timely to the users. 

Finally, the energy consumption to process task $\Omega_i$ is given by
\begin{equation}
\label{eqn_energy}
E_{i} =E_{i}^T + E_{i}^C.
\end{equation}

{\color{black}The energy consumption due to transmission,} $E_{i}^T$ is the sum of the energy consumption for transmission of the sub-tasks, and is given by
\begin{equation}
\label{eqn_transfer_energy}
E_{i,j}^T = \sum_{j=0}^K \delta_{i,j}^T P_{i,j} = \sum_{j=0}^K \frac{D_{i,j}}{\zeta_{i,j}}P_{i,j}.
\end{equation}
As shown in~\cite{7542156,8352664}, the computation energy consumption can be calculated by:
\begin{equation}
\label{eqn_transfer_energy2}
E_{i,j}^C = \sum_{j=0}^K c(f_{i,j}^k)^2C_{i,j},
\end{equation}
where $c=10^{-26}$ and $f_{i,j}^k$ is the frequency used to compute the  $j^{th}$ sub-task.

\subsection{Problem Formulation}

Theoretically, we can define the objective function as:
{\color{black}
\begin{equation}
\label{reward0}
\begin{aligned}
\max_{a_i \in \pi} & \quad \sum_i^N \beta_1\Lambda - \beta_2(E_i^T + E_i^C) - \beta_3\max(\boldsymbol{\delta}_i)\\
% & = \sum_{i=0}^N \sum_{j=0}^K \left[ \beta_1\Lambda - \beta_2\left(\frac{D_{i,j}}{\zeta_{i,j}}P_{i,j} + c(f_{i,j}^k)^2C_{i,j}\right) \+ max(\boldsymbol{\delta}_i)\right] \\
& \textrm{s.t.} \quad \delta_j^R + \delta_j^Q + \delta_{i,j}^{T}+\delta_{i,j}^{c} \leq \Delta_{\text {max }},\\
& \quad\quad f_{i,j}^k \leq f_j^{max}, f_{i,j}^k \in \boldsymbol{\mathcal{F}}_i,\\
& \quad\quad a_i = \{\boldsymbol{\Phi}_i,\boldsymbol{\mathcal{F}}_i\},\\
& \quad\quad \boldsymbol{\Phi}_i = \{\phi_0,\dots,\phi_j,\dots, \phi_K\},\\
& \quad\quad \sum_{j=0}^K \phi_j =1 ,0 \leq \phi_j \leq 1
\end{aligned}
\end{equation}

}
where $\beta_1$, $\beta_2$, and $\beta_3$ are the normalization factors. The optimization objective indicates maximizing the number of processed tasks before expiring and minimizing energy consumption. Each action contains two vectors, $\boldsymbol{\Phi}_i$ for task partitioning and $\boldsymbol{\mathcal{F}}_i$ for frequency control. Specifically, $\boldsymbol{\Phi}_i = \{\phi_0,\dots,\phi_j,\dots, \phi_K\}$ for slicing the tasks according to the edge servers, and $\phi_j$ denotes the percentage of the task offloaded to $j^{th}$ server; $\boldsymbol{\mathcal{F}}_i = \{f_0,\dots, f_j, \dots, f_K\}$ presents the recommended frequencies, and $f_j$ denotes the recommended frequency for the $j^{th}$ sub-tasks. The formulation is relatively straightforward in the mathematical definition. However, it is inflexible to balance reducing energy consumption and to increase the number of completed tasks. Therefore, we formulate the objective function so as to maximize the expected accumulated rewards given by:

\begin{equation}
\label{eqn_rewards}
\max_{a_i \in \pi} \mathbb{E} \left[ \sum_i R_i(s_i,a_i) \right]
\end{equation} 
where $s_i$ is the current system observation and $\pi$ denotes a policy; a policy maps observation states to actions. Each action $a_i$ taken by the coordinator and corresponding reward is defined as
{\color{black}
\begin{equation}
\begin{aligned}
\label{eqn_reward_time_slot}
R_i(s_i,a_i) &= \alpha w_1 \Lambda - (1-\alpha)w_2 log(E_i) \\
& - w_3 log (\max(\boldsymbol{\delta}_i)) + \mathcal{C},
\end{aligned}
\end{equation}
where $\alpha$ is the weight that allows the network providers to adjust the reward function based on their interests; $w_1,w_2$ and $w3$ are normalization terms to convert $\Lambda$, $log(E_i)$, and $log (\max(\boldsymbol{\delta}_i))$ into the same scale}, and $\mathcal{C}$ is a small incentive to encourage agents to maintain the stability if the MEC servers.

\section{D3PG: Dirichlet Deep Deterministic PolicyGradien}
\label{sec_prop_method}

In this section, we provide a {\color{black}brief introduction to DRL and introduce the developed model with extensive details.}

\subsection{Background}
The DRL settings are very similar to standard reinforcement learning. The learning {\color{black}agents interact with the environment to learn and make decisions}. For reinforcement learning, we need a description of the MEC network, which we call the environment.  {\color{black}We assume the environment represents the MEC network and provides an interface to the agent to interact with it.} In other words, reinforcement learning agents make decisions based on the observation provided by the environment, and the decisions are optimized with respect to expected long-term rewards.

{\color{black}To formulate the MEC network environment as an MDP,} we need to specify the components of the MDP, including state space, action space, and a reward denoted as $(S, A, P, R)$, and the transition function $p(s^{\prime} | s, a)$ of the MDP can be given as
\begin{equation}
\label{eqn_state_transition_prob2}
p(s^{\prime} | s, a) \doteq \operatorname{Pr}\{s_{t+1}=s^\prime | s_t=s, a_t=a \}.
\end{equation}
The transition probability function shows the transition probability of transit from current state $s$ to next state $s^{\prime}$, where $s_{t+1}$ is state of the $(t+1)^{th}$ time step; $s_t$ and $a_t$ are the current state and action of the $t^{th}$ time step assigned with current state and action, respectively. However, the transition function of the MEC network is unknown. Therefore, we prefer to design a model-free method to overcome the challenges.

The long-term accumulated rewards are also known as the return function, and it can be given as

\begin{equation}
\label{eqn_return}
G_i \doteq R_t+\gamma R_{t+1}+\gamma^{2}R_{t+2} +\cdots=\sum_{k=0}^{\infty} \gamma^{k} R_{t+k},
\end{equation}
where $R_t$ is the immediate reward and the rest of the terms denote the estimated future rewards discounted by $\gamma \in [0,1]$. {\color{black}Note that the definition of $R_i(s_i,a_i)$ is almost identical with $\beta_1\Lambda - \beta_2(E_i^T + E_i^C) - \beta_3\max(\boldsymbol{\delta}_i)$ in the target optimization problem (Eq.~\ref{reward0}); therefore, the optimization problem can be solved by maximizing long-term rewards (Eq.~\ref{eqn_return}).}

The goal of the RL models is to find optimal policies to maximize long-term expected rewards. {\color{black} A policy $\pi$ can be considered as a function mapping states to action,} and there is no policy can collect more rewards than an optimal policy $\pi^*$. Note that it is possible than we can find multiple optimal policies. To find the optimal policy $\pi^*$ that maximizes the expected long-term reward, the standard RL methods store all of the possible state and action pairs in tabular data structure, and each pair of state-action has attached a expected long-term reward. Tabular data structure called Q-table is designed to hold state-action pairs and corresponded reward values. The Q-table also can be formulated as an action-value function
\begin{equation}
\label{eqn_policy}
Q^{*}(s, a)=\max _{\pi} \mathbb{E}\left[R_\tau+\gamma G_{\tau+1} | s_\tau=s, a_\tau=a, \pi\right].
\end{equation}
Intuitively, the RL models save the learned policies, mapping states to actions, into the Q-table; the agent then searches the best actions from the table with the states. {\color{black}However, the Q-table can quickly get an explosion and it is incredibly challenging to search the optimal policies when dealing with high-dimensional or continuous state space.} Therefore, Deep Neural Networks (DNNs) have employed to capture the high-dimensional observations and generate plausible policies, which maximize expected long-term rewards.

The essential ideal of incorporating DNNs into reinforcement learning is to employ deep learning to process the complex observation and reinforcement learning to take complex actions. Although DNNs can be considered non-linear approximators to capture the high-dimensional states, DDNs are notoriously unstable in reinforcement learning because of noisy feedback and other attributes of reinforcement learning presented later of this section. Mnih \etal~\cite{dqn_nature} adopted DNNs as the approximators to capture the high-dimensional states and extract the feature maps without knowing the domain knowledge, called Deep Q-Network (DQN). Moreover, they introduce a method to delay parameter updates to stabilize the learning process. {\color{black} Specifically, they copy the network as target network, which is frozen for updating during most of the training episodes and its parameters are updated once after every $\mathcal{N}$ episode.} In DQN, the authors also adopted experience reply buffer~\cite{reply_buffer} to decouple the correlation of the sequential interaction tracks. During the learning process, the agents draw training samples from the replay buffer $U(D)$ to train the model, and Temporal Difference (TD)~\cite{td_Sutton} learning method is adopted to training the model. Specifically, the agents try to minimize the target value and current value given by the following loss
\begin{equation}
\label{eqn_loss_uniform}
\begin{aligned} 
L_\tau(\theta_\tau)=\mathbb{E}_{(s, a, r, s^\prime) \sim U(D)}
\Bigl[\Bigl(r+\gamma \max _{a^\prime} Q(s^\prime, a^\prime; \theta_\tau^-) \\ 
-Q(s, a ; \theta_\tau)\Bigr)^2 \Bigr],
\end{aligned}
\end{equation}
where $Q(s^\prime, a^\prime; \theta_\tau^-)$ and $Q(s, a ; \theta_\tau)$ denote the target and current Q-value, respectively. The $s, a ,\theta_\tau$ are the current state, action and local network weights, respectively. Similarly, $s^\prime, a^\prime, \theta_\tau^-$ are the next state, next action and target network weights, respectively. Further, the weights $\theta$ of the learning network can be updated by
\begin{equation}
\label{eqn_update_theta}
\begin{aligned}
\theta \leftarrow \theta - \eta \nabla_{\theta_\tau}L\left(\theta_\tau\right),
\end{aligned}
\end{equation}
and the target network weights are fixed and only update every $\mathcal{N}$ step by $\theta^- \leftarrow \theta$.

In this work, both the state space and action space have continuous ranges. Therefore, we have to develop a DRL model that can address continuous state space and action space; ideally, it is a policy-based model so that it does not rely on value functions to derive optimal policies. The policy-based reinforcement learning methods are easy to extend to high-dimensional or continuous action space because they do not have to reason individual actions as the objective function is parameterized policy $\pi$. In the policy-based DRL method, the models derive the optimal police directly by maximizing the expected long-term utility $\pi \approx \pi^{*}$ maximize expected return
\begin{equation}
\label{eqn_policy_based}
U(\theta)=\sum_{\tau} P(\tau ; \theta) R(\tau),
\end{equation}
where $P(\tau; \theta)$ is the probability selection actions based on the policy. The learning process (policy optimization) can be implemented with many methods such as hill-climbing, genetic algorithms, and Policy Gradient Algorithms~\cite{dpg_silver14}. The popular ways to derive the optimal policies is to use Policy Gradient as
{\color{black}
\begin{equation}
\theta \leftarrow \theta+\eta \nabla_{\theta} U(\theta)
\end{equation}
where learning rate $\eta$ decays over the time steps to avoid overshooting which an result in non-optimal policies.} Policy-based methods have many advantages such as good converge properties, easily extend to high-dimensional and continuous action space, and achieve true stochastic policy. However, policy-based RL methods also have some drawbacks. First, policy-based methods are susceptible to convergence to local optima, especially with non-linear function approximators. The issue is shared with value-based methods, but it is more challenging in the policy-based method. Second, the obtained knowledge is specific and does not always generalize well because it only captures what the agent wants to optimize the policy and includes no other information. Third, it ignores much information in the data.

\subsection{Proposed Model}

\begin{figure*}
  \centering
  \includegraphics[width=7in]{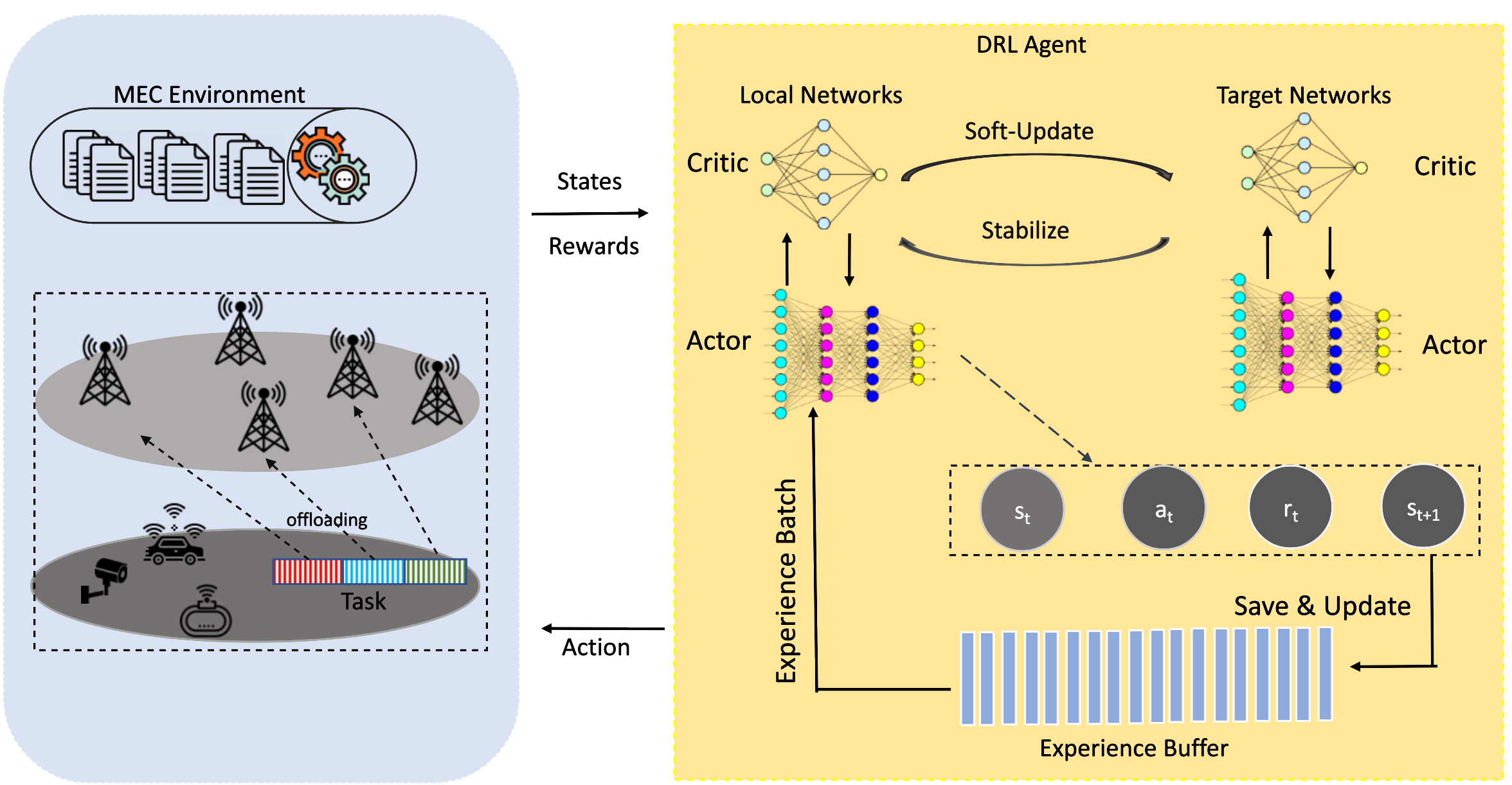}
  \caption{Developed Model}
  \label{fig:arch_model}
\end{figure*}

Considering the advantages and disadvantages of the value-based and policy-based DRL, we prefer to develop an Actor-Critic DRL model because it has the advantages of value-based and policy-based reinforcement learning. Specifically, we develop the model called Dirichlet Deep Deterministic Policy Gradient (D3PG), which builds on Deep Deterministic Policy Gradient (DDPG)~\cite{Lillicrap2016}. The developed model has to address continuous action space and meet the constraints of MEC task partitioning. Specifically, a task can be partitioned as $K$ sub-tasks, and the size of sub-tasks can be represented by $\boldsymbol{\Phi}_i = \{\phi_1,\dots,\phi_j,\dots, \phi_K\}$,  $\phi_j$ denotes what percentage of the full task is contained in the $j^{th}$ sub-task. Thus the sum is constrained by $\sum \Phi_i = 1$, since the total percentage is one. {\color{black}To satisfy action space constraints, we employ the Dirichlet distribution to capture the constrained action.   

Fig.~\ref{fig:arch_model} shows the work process of task partitioning and offloading as well as frequency control with the developed DRL model.} The system has three components: the MEC network, the MEC environment, and the DRL agent. The learning DRL agent has no control over the MEC network directly; instead, the environment serves as a coordinator to bridge the MEC network and the DRL learning agent. We assume the environment presents the MEC network for the simplicity of argument, and the DRL interacts with the environment and learns from the trials.

The learning process can be considered as the following steps. First, the agent takes action given the observation from the environment. Second, the environment provides feedback and the next state to the DRL agent; {\color{black}the agent then stores the current interaction data into the experience replay buffer for training the model.} Each record of interaction includes current state, action, reward, and next states denoted as a tuple $<s_t,a_t,r_t,s_{t+1}>$. The DRL agent keeps interacting with the environment to generate the training data sets. Third, the agent then draws training data from the experience replay buffer to train the learning networks inside the DRL model. Each network has a backup copy called target network, and the target networks are for stabilizing the training. The details of the elements of the MDP and DRL training process are presented in the following subsections.  

\subsubsection{Action}

Although the DDPG models can address many DRL challenges with continuous action space, part of the action space in this work has a further constraint. Specifically, each action has two vectors, one vector $\boldsymbol{\Phi}_i = \{\phi_1,\dots,\phi_j,\dots, \phi_n\}$ for {\color{black} partitioning the tasks into sub-tasks according to the edge servers,} and the other vector $\boldsymbol{\mathcal{F}}_i = \{f_0,\dots, f_j, \dots, f_n\}$, where $n$ is the number of edge servers. In other words, the DRL specifies $p_j$ percentage of the task to offloads to the $j^{th}$ edge server; further, it recommends the server process the sub-task by using $f_j$ percentage of maximum CPU frequency of the $j^{th}$ edge server. All of the elements (sub-component of the action) are continuous, and in the range $[0,1]$. {\color{black}In other words, we can define the constraints as $\phi_j \in [0,1]$ and $f_j \in [0,1]$, and the $j^{th}$ edge server should not receive a sub-task when  $\phi_j =0$.} Moreover, the sum of proportion of the sliced tasks has to equal 1. Therefore, a specific action $a_i$ can be given,  
{\color{black}
\begin{equation}
\begin{aligned}
\label{eqn_action_form}
a_i &= \{\boldsymbol{\Phi}_i,\boldsymbol{\mathcal{F}}_i\}\\ 
& \textrm{s.t.} \quad 0 \le \phi_j \le 1, \sum_j^n \phi_j = 1.
\end{aligned}
\end{equation}

Indeed, the $softmax$ function can satisfy the constraint that $\sum_j^n \phi_j = 1 $. However, the $softmax$ function does not have an exploration mechanism, which probably leads the model to a local optimum.} In DRL, the learning agent has to explore the environment because the feedback (rewards and punishments) is not labeled data as supervised machine learning; it is an evaluation score of the actions and policies. Therefore, we cannot regard the feedback as the actual label data as the deep Learning training. A possible way to mitigate this problem is to add an exploration method such as $\epsilon-greedy$ by drawing $\boldsymbol{\Phi}_i$ from a random distribution (e.g., uniform distribution) or add a noise vector for each action. However, those methods are unstable to explore the continuous action space. Therefore, we use a Dirichlet distribution to characterize the $\boldsymbol{\Phi}_i$, which means $\boldsymbol{\Phi}_i \sim Dir(\phi)$. Dirichlet distribution can not only satisfy the constraint of the $\boldsymbol{\Phi}_i$ but naturally explores the possible actions to find the optimal policies by sampling from the Dirichlet distribution. Given the random process of Dirichlet sampling, the agent has achieved stochastic policy without saving specific actions. Therefore, the $\boldsymbol{\Phi}_i \sim Dir(\phi)$ is defined as
{\color{black}
\begin{equation}
\begin{aligned}
\label{eqn_action_dirch}
Dir(\phi) &= \frac{1}{B(\boldsymbol{\Psi}_i)} \prod_{j=1}^n \phi_j^{\psi_j-1}, \phi_j \ge 0,\\
& where \
B(\boldsymbol{\Psi}_i) = \frac{\prod_{j=1}^K \Gamma(\psi_j)}{\Gamma(\sum_{j=1}^K \psi_j), \psi_j>0}
\end{aligned}
\end{equation}
where $\Gamma(\cdot)$ is a standard Gamma function given by
\begin{equation}
\Gamma(z)=\int_{0}^{\infty} x^{z-1} e^{-x} d x, \quad \Re(z)>0
\end{equation}

\begin{figure}[!t]
  \centering
  \includegraphics[width=4.2in]{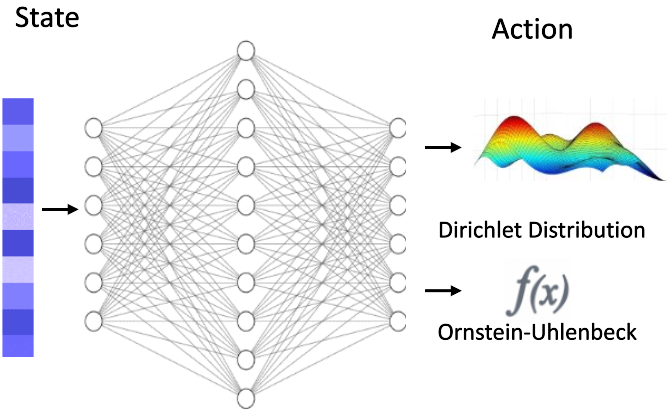}
  \caption{Actor Network.}
  \label{fig:actor}
\end{figure}

Moreover, to meet the condition $\psi_j > 0$  of the Dirichlet function, we use an exponential to process the actor-network outputs for the slicing action $\boldsymbol{\Psi}_i = e^{\boldsymbol{z}} + \epsilon$}, where $\boldsymbol{z}$ is post-activation (outputs) parts of the previous layer, and $\epsilon$ is a very small positive number to avoid zero values. {\color{black}Theoretically, $e^z$ is always greater than 0; however, it could be extremely close to 0 in some cases which would raise issues in the real-wold implementation, and $\epsilon$ is to avoid such implementation errors. } The other parts, post-activation from the previous layer, are inputted to the Ornstein-Uhlenbeck process. As shown in Fig.~\ref{fig:actor}, the results of the Dirichlet distribution and Ornstein-Uhlenbeck process are concatenated as a complete action which can be given, 
\begin{equation}
\mu^{\prime}\left(s_{t}\right)= Dir(\theta_{t}^\phi) \boldsymbol{\oplus} \left[\mu\left(s_{t} \mid \theta_{t}^f \right)+\mathcal{N}\right]
\end{equation}
where $\boldsymbol{\oplus}$ denotes the concatenation of two vectors, and $Dir(\phi_{t}^\phi)$ sub-actions for task partitioning, and the rest of the elements are for the frequency control. The Dirichlet distribution and Ornstein-Uhlenbeck process, denoted $\mu\left(s_{t} \mid \theta_{t}^f \right)+\mathcal{N}$, can address the exploration challenges during the learning phase. Therefore, the developed model can keep exploring the environment and is unlikely stacked to non-optimal policies because the DRL model keeps exploring by sampling actions from Dirichlet distribution and adding noise to actions with Ornstein-Uhlenbeck process.

\subsubsection{State Space}
Although the MEC environment is not fully observable to the DRL agent, which means the observation is not equivalent to the state, {\color{black}we assume the observation is the same as the other reinforcement learning methods. The state space at time slot $t$ is denoted as $s_t = <\mathcal{M},\boldsymbol{\zeta}, \boldsymbol{\Omega}$, where $\mathcal{M}$ denotes the status of MEC servers, $\boldsymbol{\zeta}$ is transmission rate matrix, and $\boldsymbol{\Omega}$ 
is the set of the current tasks ready to offload}.  Each of the components of the state has its own sub-components as defined in section.\ref{sec_sys_model}.

\subsubsection{Transition Probability and Reward Function}
As the standard reinforcement learning methods, we assume the environment fits in MDP. However, we do not know the transition function (Eq.~\ref{eqn_state_transition_prob2}) in the environment. Finally, the reward function for a specific time step is defined by Eq.~\ref{eqn_reward_time_slot}.

\subsection{Loss Function and Gradient}

Constructing the loss function is one of the critical steps of the training DRL models, and we need two loss functions: one for the actor-network and another for the critic-network. However, it is unnecessary to provide an explicit form loss function for the actor-network because the actor-network is optimized with respect to the critic value. The actor-network is policy-based, and the critic network is value-based; therefore, we can consider the critic as the Q-Network as in the DQN model~\cite{dqn_nature}. The critical value is the utility for the policy-based actor-network. The learning process of the critic is very similar to Q-learning and DQN. The training process of Q-learning uses Temporal Difference (TD)~\cite{td_Sutton} to update the Q-values so that the agents can search policies from the Q-table. The Q-value $Q(s, a)$ updating can be accomplished with

\begin{equation}
\begin{aligned}
& Q(s, a) \leftarrow Q(s, a)+\eta \delta_t,\\
&\delta_t = \Bigl[r_{t} +\gamma \max _{a} Q\left(s^{\prime}, a^{\prime}\right)-Q(s, a)\Bigr],
\end{aligned}
\end{equation}
where $\eta$ is the learning rate, $\delta_t$ is the TD error, and $\gamma \in [0,1]$ is the discount factor of the expected feature values. Theoretically, it has been proven~\cite{Watkins1992,SuttonRL} that the near optimal Q-value can obtained by iterating the above steps until $|Q^\prime(s,a) - Q(s,a)| < \epsilon$, where $\epsilon$ is a very small positive number. 

Similarly, the action-value function of actor-critic can be given as the Bellman equation
\begin{equation}
\begin{aligned}
\label{eq_q_value}
Q^{\pi}\left(s_{t}, a_{t}\right) &=\mathbb{E}_{r_{t}, s_{t+1} \sim E}\Bigl[r\left(s_{t}, a_{t}\right) \\
&+\gamma \mathbb{E}_{a_{t+1} \sim \pi}\left[Q^{\pi}\left(s_{t+1}, a_{t+1}\right)\right]\Bigr],
\end{aligned}
\end{equation}
where $r(s_t, a_t)$ is the immediate reward when the agent takes action $a_t$ based on the given state $s_t$, and the remaining terms are estimated future values based on policy $\pi$ discounted by $\gamma$. In actor-critic DRL, the actions are made by the actor-network, which is parameterized by $\mu(s|\theta)$. Assuming the policies are deterministic, we can derive the Q-value function when the actions are generated by an actor-network $\mu\left(s_{t+1}\right)$ given by 
\begin{equation}
\begin{aligned}
Q^{\mu}\left(s_{t}, a_{t}\right) &=\mathbb{E}_{r_{t}, s_{t+1} \sim E}\Bigl[r\left(s_{t}, a_{t}\right) +\gamma Q^{\mu}\left(s_{t+1}, \mu\left(s_{t+1}\right)\right)\Bigr].
\end{aligned}
\end{equation}
Therefore, the loss function based on Q-learning~\cite{Watkins1992} or DQN~\cite{dqn_nature}, and the targeted minimize function can be given,
\begin{equation}
\begin{aligned}
L\left(\theta^{Q}\right)=\mathbb{E}_{s_{t} \sim , a_{t} \sim \beta, r_{t} \sim E}\left[\left(Q\left(s_{t}, a_{t} \mid \theta^{Q}\right)-y_{t}\right)^{2}\right],
\\ 
s.t.\quad y_{t}=r\left(s_{t}, a_{t}\right)+\gamma Q\left(s_{t+1}, \mu\left(s_{t+1}\right) \mid \theta^{Q}\right),
\end{aligned}
\end{equation}
where $\rho^{\beta}$ is the state transition probability given the action distribution $\beta$. 
Note that $y_t$ is parameterized by the actor-network $\mu(s_{t+1})$. Further, the gradient of the loss function can be derived with the chain rule as

\begin{equation}
\begin{aligned}
\label{eq_graident_loss}
& \nabla_{\theta^{\mu}} J \approx \mathbb{E}_{s_{t} \sim \rho^{\beta}}\left[\left.\nabla_{\theta^{\mu}} Q\left(s, a \mid \theta^{Q}\right)\right|_{s=s_{t}, a=\mu\left(s_{t} \mid \theta^{\mu}\right)}\right] \\
&=\mathbb{E}_{s_t \sim \rho^\beta}\left[\nabla_{a} Q \left(s, a \mid \theta^Q\right)|_{s=s_t, a=\mu(s_t)} \times \nabla_{\theta_\mu} \mu\left(s \mid \theta^\mu \right) \mid_{s=s_t}\right]
\end{aligned}
\end{equation}
To stabilize the training process, DDPG also requires the target networks to compute temporal differences. Therefore, $y_t$ is obtained from the the target critic-network $\theta^{Q^\prime}$ and actor-network $\theta^{\mu^\prime}$. The final loss function can be given
\begin{equation}
\begin{aligned}
L\left(\theta^{Q}\right)&=\mathbb{E}_{s_{t} \sim \rho^{\beta}, a_{t} \sim \beta, r_{t} \sim E}\left[\left(Q\left(s_{t}, a_{t} \mid \theta^{Q}\right)-y_{t}\right)^{2}\right]\\
&y_{t}=r\left(s_{t}, a_{t}\right)+\gamma Q\left(s_{t+1}, \mu^\prime\left(s_{t+1}\right) \mid \theta^{Q^\prime}\right),
\end{aligned}
\end{equation}
where $y_t$ is the Q-value computed by the target critic-network. 

\subsection{Training process}
The training process is shown in Alg.~\ref{agl:model_train}. The algorithm has three blocks, including initialization, data generation and collection, and model training.

\begin{algorithm}
  \caption{D3PG for Task Partitioning and Offloading and Frequency Control\label{agl:model_train}}
  \KwIn{$epoch\_no$, $\mathcal{M}$}
  \KwOut{$loss, gradients$}
  //1. Initialization: \\
  Initialize replay memory $\mathcal{D}$; \\
  Randomly initialize $\theta^Q$ and $\theta^\mu$;\\
  Initialize actor-network $\mu(s \mid \theta^Q)$ with weights $\theta^\mu$; \\ 
  Initialize critic-network $Q(s,a \mid \theta^Q)$ \\
  Initialize target networks $Q'$ and $\mu'$ weights: $\theta^{Q'} \leftarrow \theta^Q$, $\theta^{\mu'} \leftarrow \theta^\mu$; \\
  
  \For{$episode \leftarrow 1\ to\ \mathcal{M}$}
  {
    % Reset the environment;\\
    Initialize a random process $\mathcal{N}$ for action exploration;\\
    Preprocess initial state: $S \leftarrow \psi(<x_1>)$;\\
      \For{time step: $\tau \leftarrow 1\ to \ T_{max}$}
      {
        // 2. Generate training data:\\
        Select action according to the current policy and exploration noise:\\
        $a_t = Dir(\theta_{t}^\phi) \boldsymbol{\oplus} \left[\mu\left(s_{t} \mid \theta_{t}^f \right)+\mathcal{N}_t\right]$; \\ 
    
        Execute action $a_t$ and observe reward $r_t$ and next state $S^\prime$;\\
        Store experience $(S,A,R,S^\prime)$ in $\mathcal{D}$; \\
        // 3. Learning: \\ 
        Obtain random mini-batch of $(s_i,a_i,r_i,s_{i+1})$ from $\mathcal{D}$;\\
        $y_i = r_i + \gamma Q_i^\prime(s_{i+1}, \mu^\prime(s_{i+1} \mid \theta^{\mu^\prime}))$;\\
        Update critic by minimizing the loss:\\
        $L(\theta) = \frac{1}{N} \sum_i(y_i - Q(s,a \mid \theta_i^Q))^2$; \\
        Update critics: $\theta_i \leftarrow \min_{\theta_i} L(\theta)$;\\
   
        Update the actor policy using the sampled policy gradient:\\
        $\nabla_{\theta^\mu} J \approx
        \frac{1}{N} \nabla_{a} Q\left(s, a \mid \theta^{Q}\right) \nabla_{\theta_{\mu}} \mu\left(s \mid \theta^{\mu}\right)$\\
        Update the target critic network:\\
        $\theta ^{Q^\prime} \leftarrow \tau \theta^Q + (1-\tau)\theta^{Q^\prime}$;\\
        Update the target actor network:\\
        $\theta ^{\mu^\prime} \leftarrow \tau \theta^\mu + (1-\tau)\theta ^{\mu^\prime}$;
        
      }
  }
\end{algorithm}
The first block is to initialize variables and the networks with random weights, creating an experience reply buffer, copying the networks to the target network. As mentioned before, we have two networks, the actor-network and the critic-network, and each network has a target network to stabilize the training. The experience reply buffer maintains the training data collected from the interaction with the MEC network environment.

The second block of the algorithm is to collect data by interacting with the environment. As mentioned earlier, the action consists of a Dirichlet distribution and Ornstein-Uhlenbeck process. Every interaction with the MEC generates a training sample, and each sample includes the current observation state, reward(feedback), next state, and a termination flag. The collected datasets are stored in the experience reply buffer, which is a queue-like container. The experience reply buffer has a fixed size, {\color{black} and it discards the oldest data when it receives new data.}

The third block is for training the networks in the model. During the training, the target policy actor has added a smooth factor $\epsilon$ with clipped range. Again, the noise is only added to the frequency control sub-actions because the rest of the sub-actions are sampled from the Dirichlet distribution. As the standard actor-critic setting, the policy is optimized with respect to the Q-values defined by the critics. Further, the target networks are updated with soft-update and delay update methods. The target network updates are delayed to reduce variance. This method is similar to the fixation method introduced in DQN; the only difference is that it updates the network more frequently than the fixation method. The soft-updates keep a significant amount of the original weights instead of completely overwriting the networks so that the model does not have to wait for a long time to update the networks to avoid high variance. The portion of weights updated to the target networks can control the factor $\tau$. 

\section{Simulation Results}
\label{sec_sim_res}

In this section, we present the details of the simulation and results analysis. We adopt {\color{black}Numpy~\cite{Charles2020} as a tool for data preprocessing} and Pytorch\footnote{https://pytorch.org/} to build the DRL models. Again, we consider the simulation has two parts, including the MEC network environment and the DRL model. The MEC network has various network entities such as edge servers and IoTs users. As we consider the MEC networks are heterogeneous, and the edge servers are configured with different computational resources; and the IoTs users frequently generate various tasks to offload. In addition, the MEC network also maintains the network properties such as the channel gain and transmit speed matrices. The entities of the MEC networks are simulated with processors. To verify our model, we compared our model (D3PG) with the existing methods, including DDPG, DDPG with $softmax$ (DDPG-softmax), Twin Delayed Deep Deterministic Policy Gradient (TD3)~\cite{Fujimoto2018} and greedy algorithm. The DRL models are implemented with Pytorch. The key parameter settings are summarized in Table.~\ref{tabke:para}.

\begin{table}[h]

\centering

\caption{Parameter Settings}
\label{tabke:para}
\begin{tabular}{p{4.5cm}|p{4.5cm} } 
 \hline
  \hline
  \bf Parameter & \bf Value \\ \hline

Signal to Noise Ratio (dB) & 100 \\ 
 \hline

Task Data Size (bits)& $[2 \times 10^5,2 \times 10^7]$\\ 
 \hline
Task Computing Size (cycles) & $[8 \times 10^6, 1\times10^7]$ \\ 
 \hline 
Server Max Frequency (Hz) & $[2\times10^9,8\times10^9]$ \\ 
 \hline 
Number of Online Users & $[10,1000]$ \\ 
 \hline  
Number of Edge Server s& $[5,50]$ \\ 
 \hline 
Batch Size & 256\\ 
 \hline 
Learning Rate $\alpha$ & $5\times 10^{-4}$\\
 \hline 
Discount Factor $\gamma$ & 0.9\\ 
 \hline  
 \hline
\end{tabular}

\end{table}

As mentioned in previous sections, both the DDPG and TD3 models have two types of neural networks, {\color{black}one  for taking the actions called the actor-network and the other for evaluating the actor-network called critic-networks.} For both models, the number of layers and the number of neurons in the hidden layer are the same. Specifically, the actor-networks have five layers, and the number of neurons is the size of the state space, 256, 512, 256, and the size of the action space, respectively. Similarly, the critic-networks have five layers, and the number of neurons in the state space plus the size of the action space, 256, 512, 256, and 1, respectively. Note that the TD3 model has two critic-networks while the DDPG has a single critic-network only. Therefore, TD3 consumes more computational power than DDPG models.

\begin{figure}
  \centering
  \begin{minipage}[b]{0.4\textwidth}
  \includegraphics[width=3.0in]{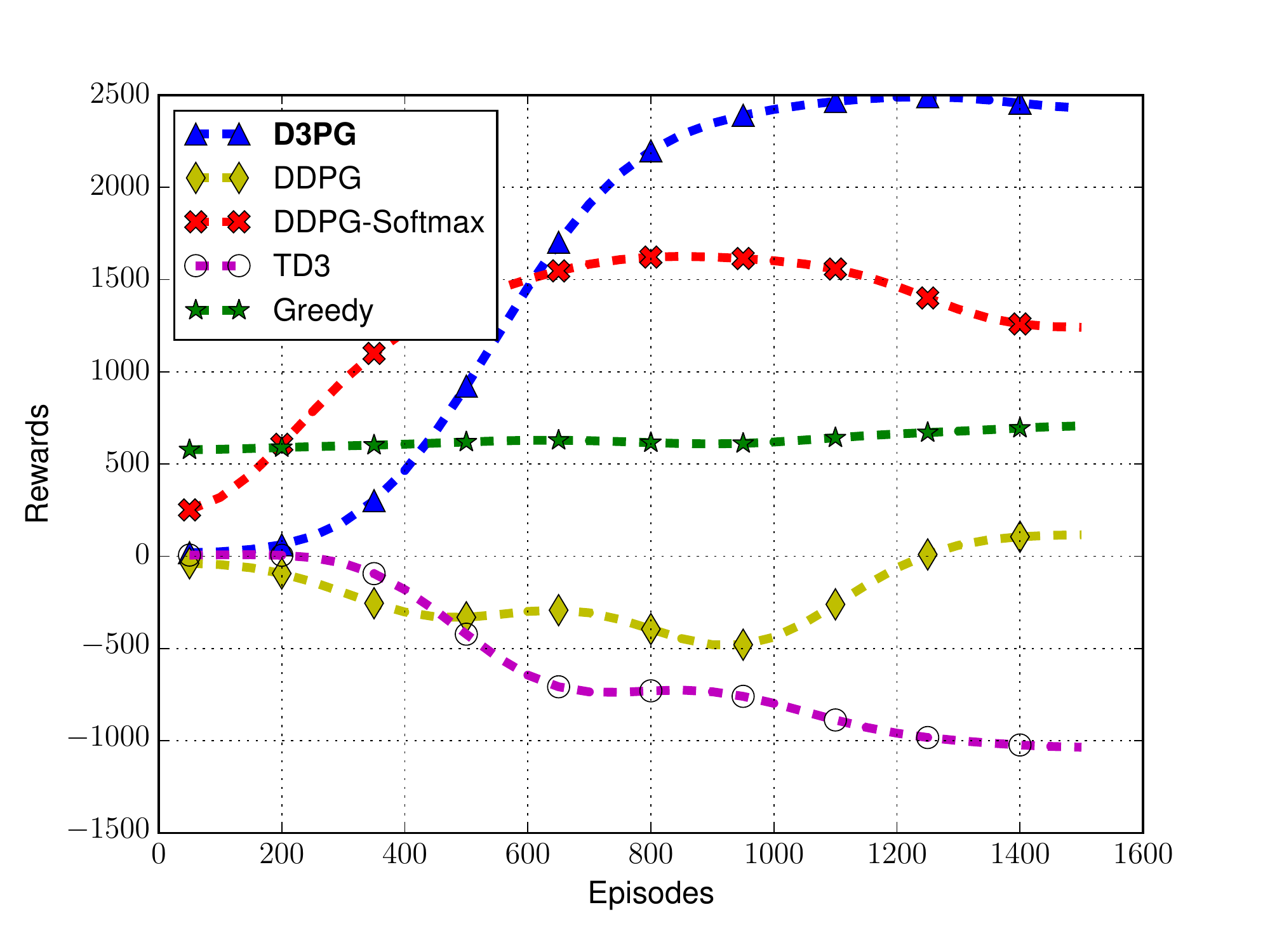}
  \caption{Reward}
  \label{fig:rewards}
  \end{minipage}
  \hfill
  \begin{minipage}[b]{0.4\textwidth}
  \centering
  \includegraphics[width=3.0in]{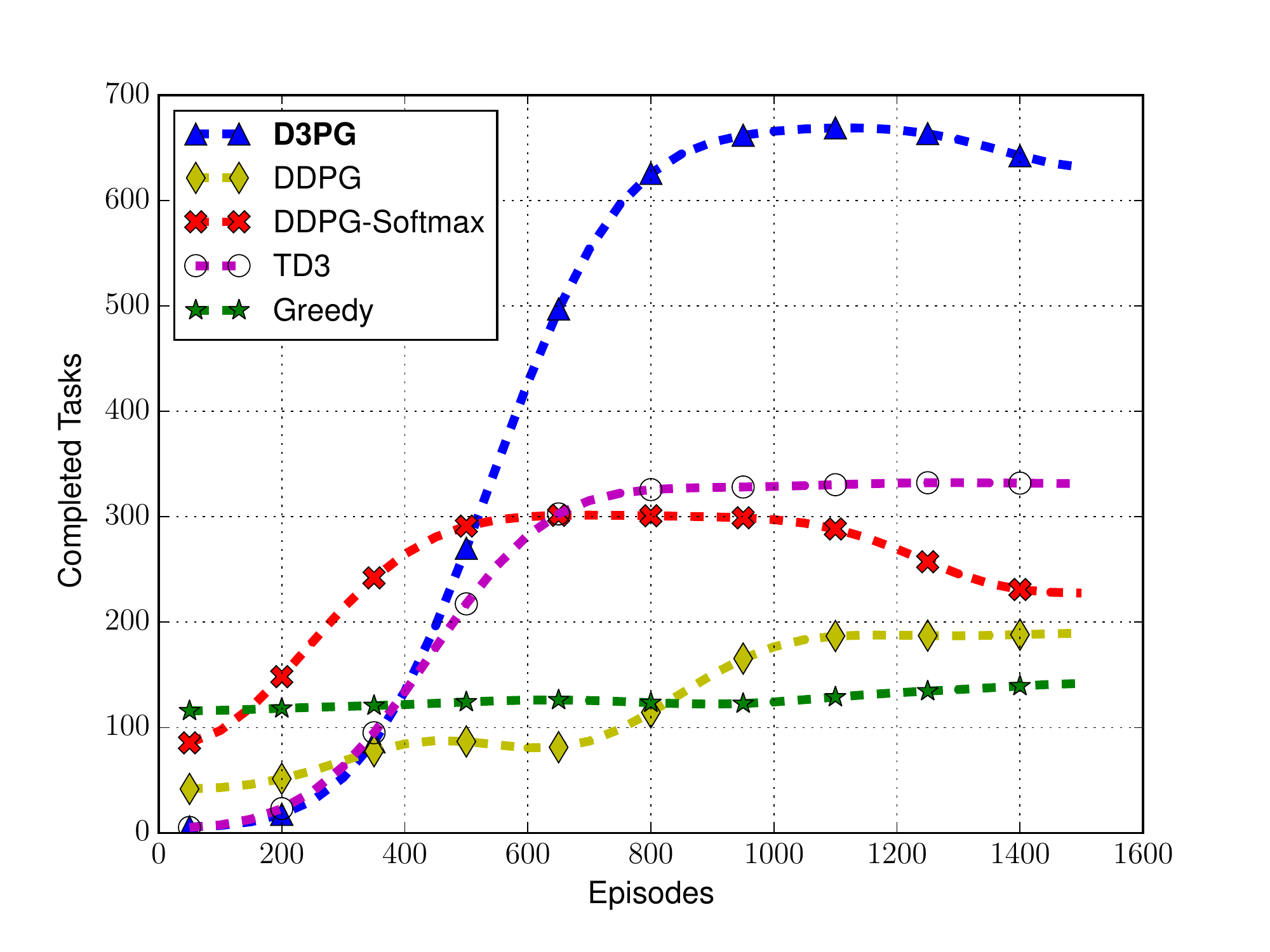}
  \caption{Completed Tasks}
  \label{fig:task}
  \end{minipage}
\end{figure}

Considering the randomness in the MEC network and DRL models, we have run five experiments and average the results. Fig.~\ref{fig:rewards} shows the rewards with respect to the episodes, and the D3PG are converged to the optimal policies around 1,500 episodes. Each reward has three components: the completed number of tasks, energy consumption, and time cost; the weights of components allow the network providers to configure based on their applications and business purposes. As we can see from Fig.~\ref{fig:rewards}, the D3PG model can achieve better results than other models because the Dirichlet distribution captures partitioning actions to improve the policies. DDPG-softmax outperforms the original version of DDPG and TD3 because $softmax$ can capture partitioning actions; however, it is highly likely to converge local optima because $softmax$ does not have an exploration mechanism to explore the optimal policies. In fact, this DDPG-softmax has relatively good results because we add noise to actions as in TD3 to help $softmax$ explore the partition actions. Both original DDPG and TD3 perform poorly because we have to force the partition actions to satisfy {\color{black}the action space constraint $\sum_j^n \phi_j = 1$, which can degrade the overall performance. The greedy algorithm does not require a learning process and can collect more rewards at early episodes; it outperforms the standard TD3 and DDGP. The D3PG and DDPG-softmax can adequately address the action space to maximize the accumulation of long-term rewards and accumulate more rewards at later episodes.}

% \begin{figure}
%   \centering
%   \includegraphics[width=3.5in]{figures/mean_scores.pdf}
%   \caption{Reward}
%   \label{fig:rewards}
% \end{figure}

\begin{figure}
  \centering
  \begin{minipage}[b]{0.4\textwidth}
  \includegraphics[width=3.0in]{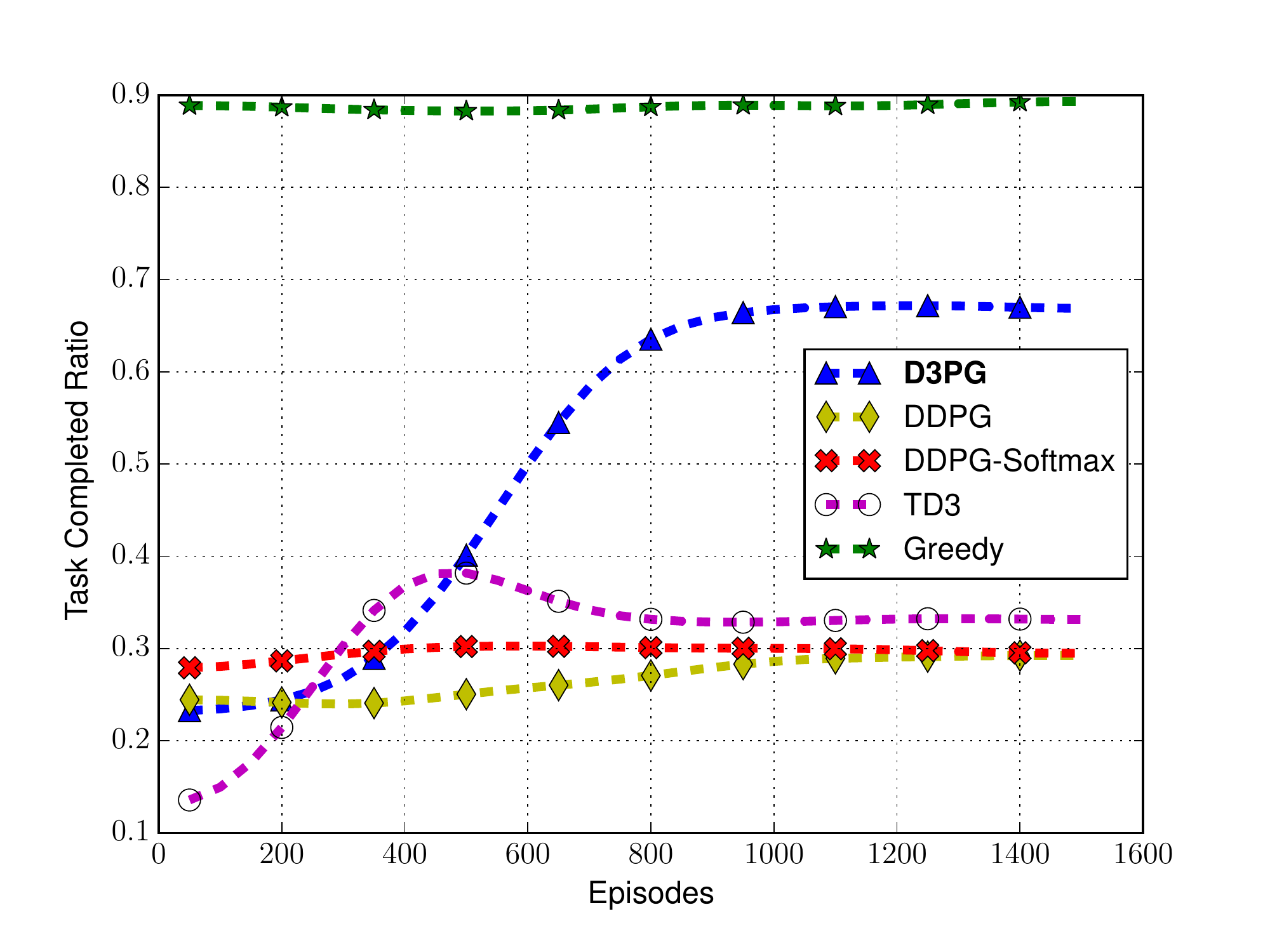}
  \caption{Task Completed Ratio}
  \label{fig:mean_task}
  \end{minipage}
  \hfill
  \begin{minipage}[b]{0.4\textwidth}
  \centering
  \includegraphics[width=3.0in]{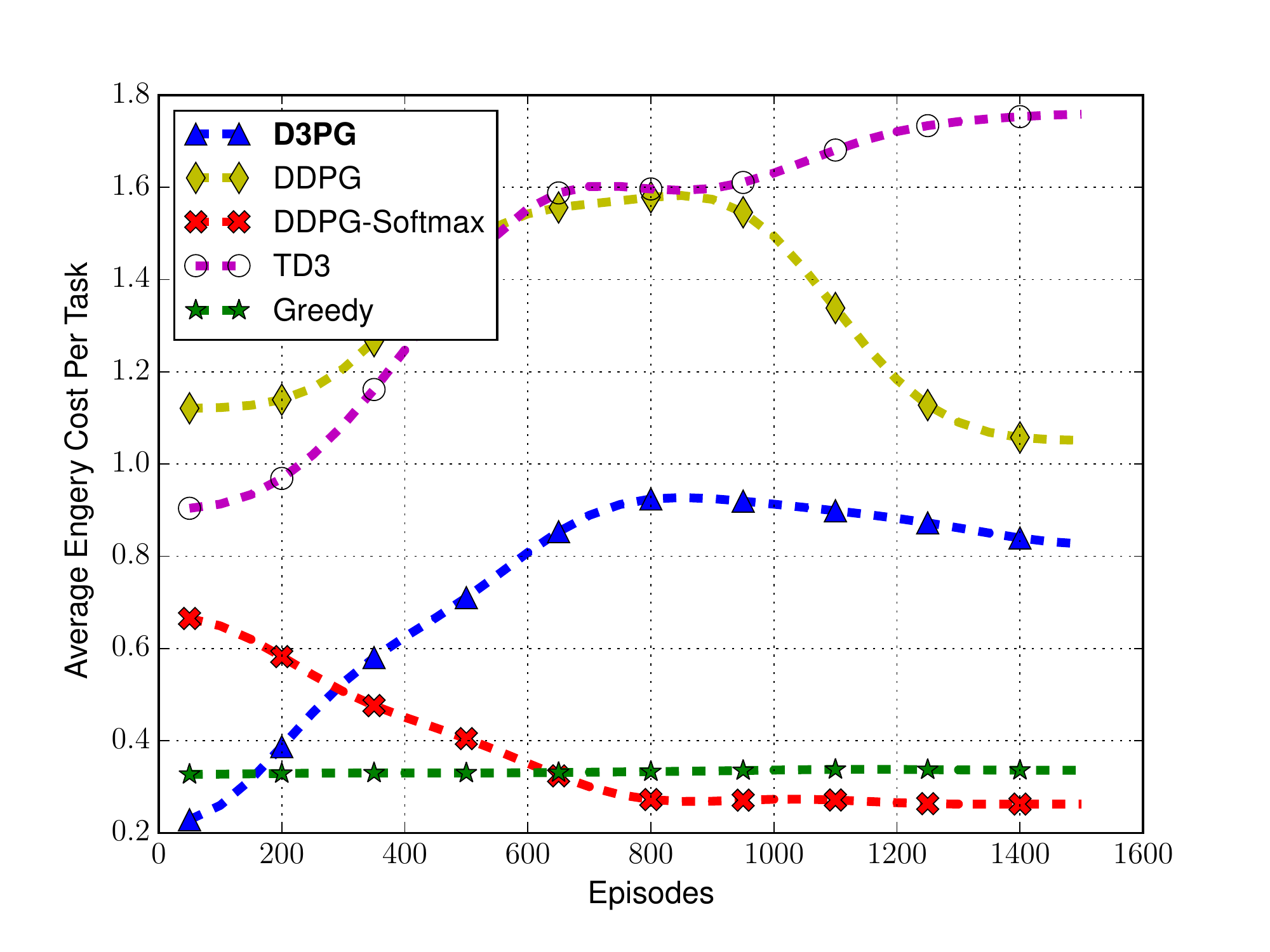}
  \caption{Energy Consumption}
  \label{fig:energy}
  \end{minipage}
\end{figure}

Note that Fig.~\ref{fig:rewards} shows the TD3 model accumulates negative rewards and performs poorer than other models. However, it learns to maximize the completed number of tasks before expiring, as shown in Fig.~\ref{fig:task}. Although the DRL model is learning to maximize the expected long-term rewards, we can decompose the rewards to verify that the model can address joint optimization in an end-to-end manner. Fig.~\ref{fig:task} shows the completed number of tasks before expiring. The edge servers can only process a minimal number of tasks at each episode at the beginning because the models take random actions that fail to allocate the resources properly. As the models interact with the MEC environment to learn and improve the policies, they can allocate the resource optimally and serve a maximum number of offloaded tasks. Fig.~\ref{fig:mean_task} shows the ratio of the number of tasks processed before expiring to the total tasks. Note that we set the computational cost and data size of tasks considerably large for the edge servers. {\color{black}Therefore, some of the tasks are even impossible to be completed before their corresponding deadlines, and the completion ratio is only for comparison purposes. The completion ratio is much higher if we reduce the size of the tasks and the computational demand. Although the greedy algorithm can collect more rewards than DDPG and TD3, the completed number of the tasks is less than the learning methods. The learning agents need to stabilize the environment and process tasks as many as possible to collect more rewards in an episode. The D3PG outperforms other methods in terms of completed tasks and task completed ratio because the constraints do not cripple the model. Moreover, the Dirichlet distribution can capture the uncertainty of the environment and explore optimal policies.}

% \begin{figure}
%   \centering
%   \includegraphics[width=3.5in]{figures/sum_task.pdf}
%   \caption{Completed Tasks}
%   \label{fig:task}
% \end{figure}

% \begin{figure}
%   \centering
%   \includegraphics[width=3.5in]{figures/mean_task.pdf}
%   \caption{Task Completed Ratio}
%   \label{fig:mean_task}
% \end{figure}

\begin{figure}
  \centering
  \begin{minipage}[b]{0.4\textwidth}
  \includegraphics[width=3.0in]{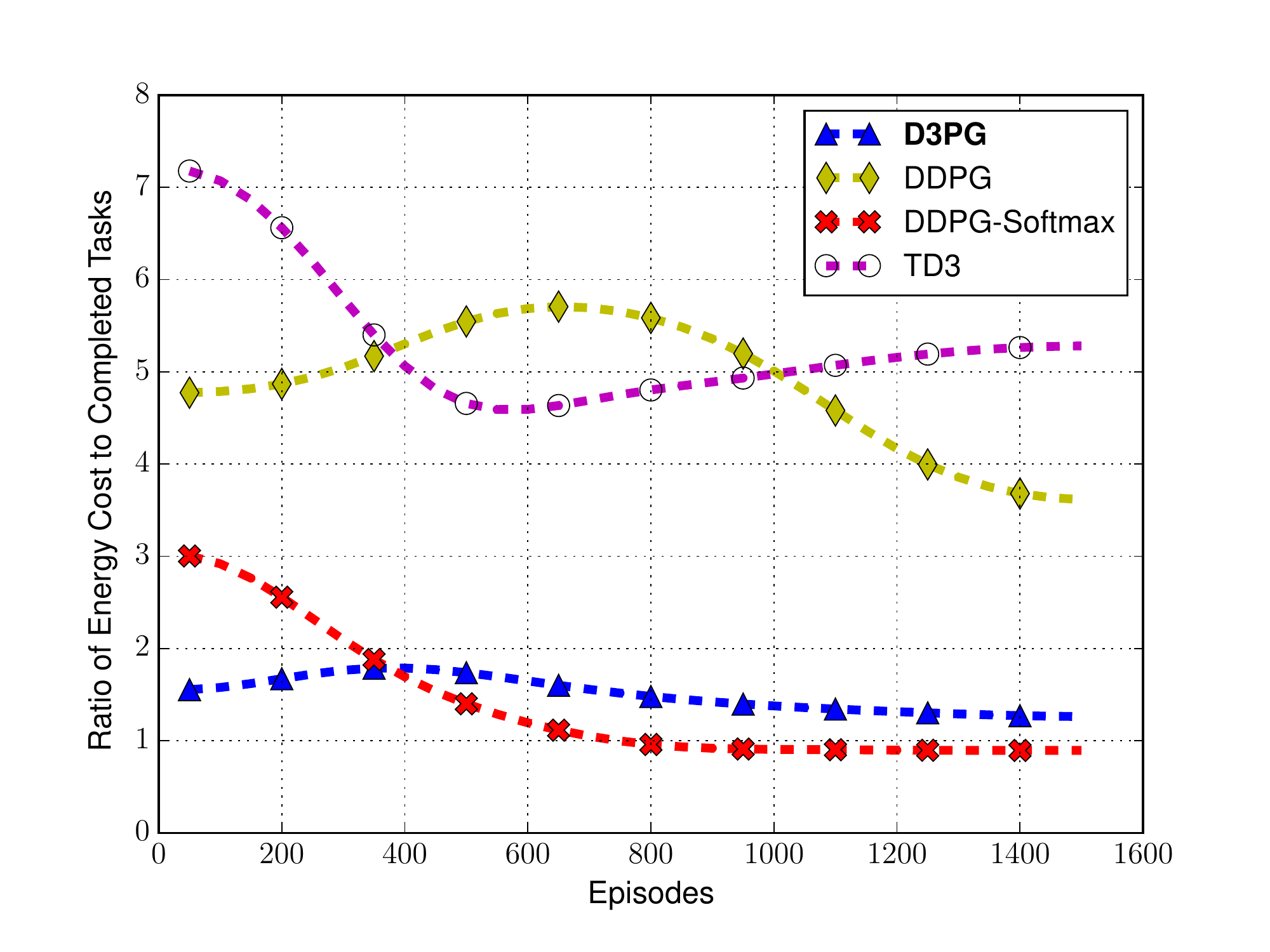}
  \caption{Energy to Tasks}
  \label{fig:energy_to_task}
  \end{minipage}
  \hfill
  \begin{minipage}[b]{0.4\textwidth}
  \centering
  \includegraphics[width=3.0in]{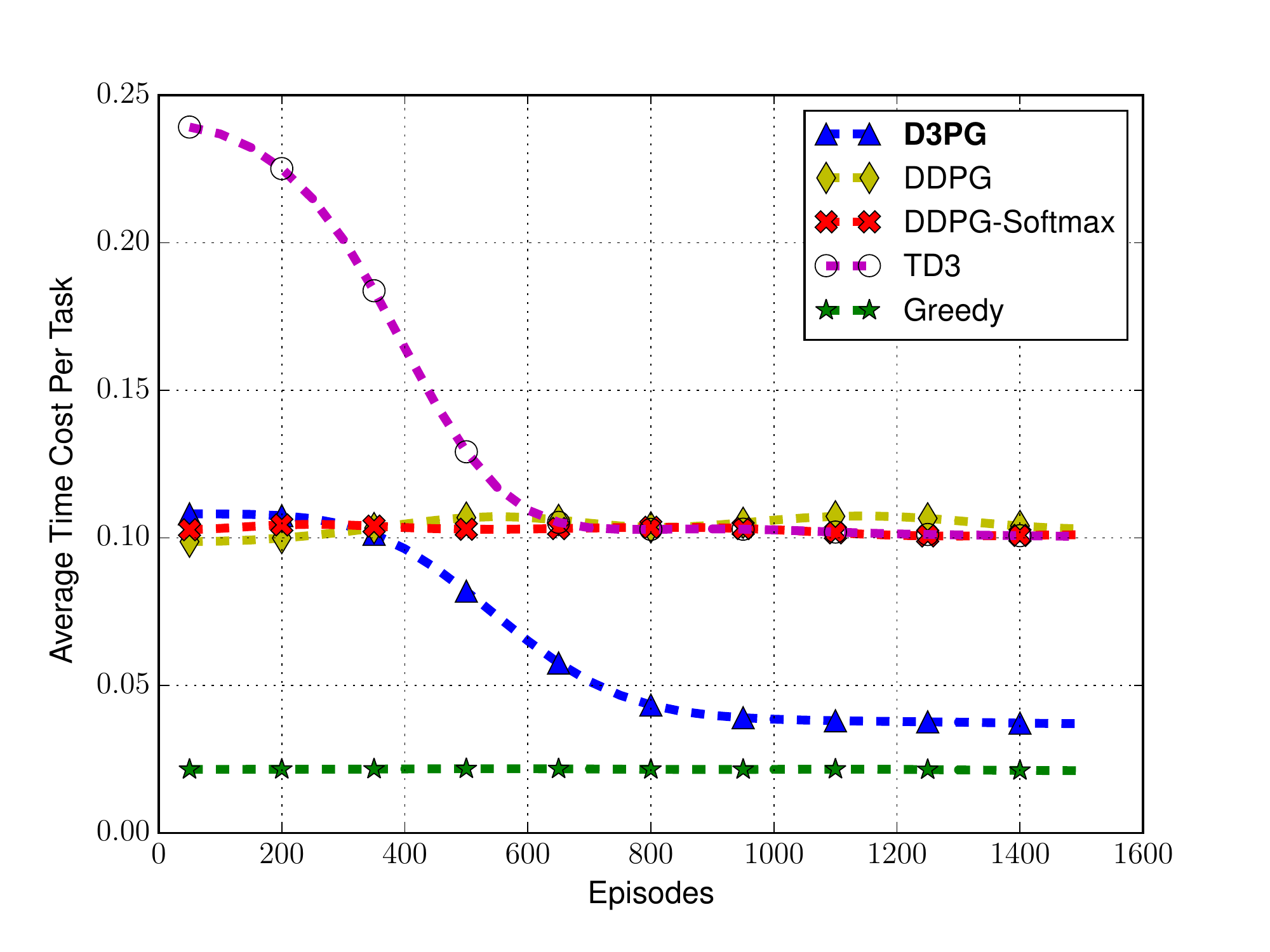}
  \caption{Average Time Cost}
     \label{fig:time}
  \end{minipage}
\end{figure}

\begin{wrapfigure}{r}{0.5\textwidth}
  \centering
  \includegraphics[width=3.5in]{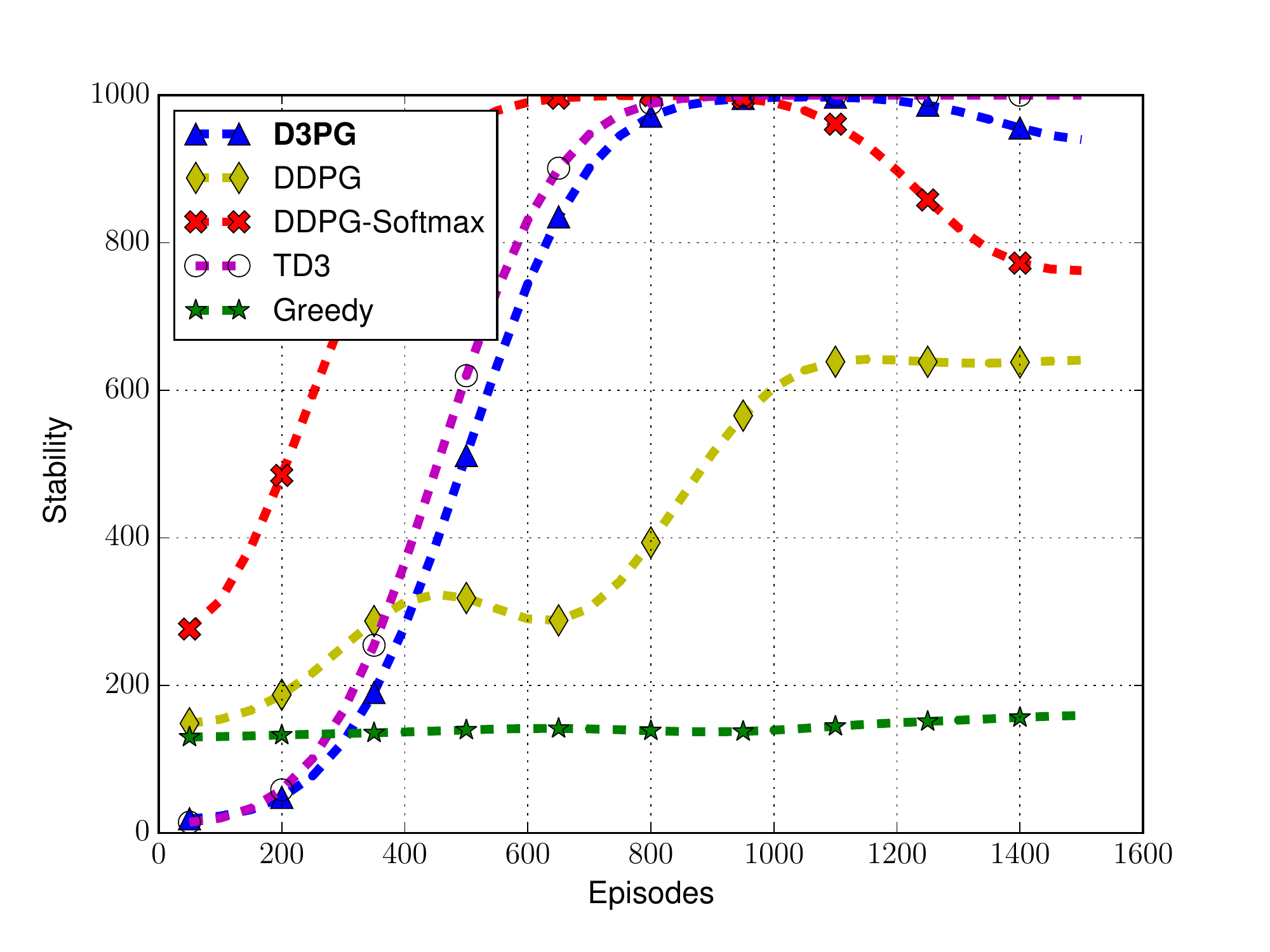}
  \caption{Stability}
  \label{fig:stable}
\end{wrapfigure}

Similarly, the models can reduce the energy consumption while maintaining the number of completed tasks, as shown in Fig.~\ref{fig:energy}. The models can save energy through frequency control because the energy consumption is proportional to the square of the CPU spinning frequencies. Therefore, the DRL models can find optimal frequencies to process the offloaded tasks to balance completed tasks and energy consumption. The TD3 consumes more energy than the rest of the models, and that explains why TD3 collects less rewards shown in Fig.~\ref{fig:rewards}. Fig.~\ref{fig:energy} shows the energy cost per task, and the get expired tasks also consume energy. We can compute wasted energy by multiple the energy consumption with the task completed ratio shown in Fig.~\ref{fig:mean_task}. Note that the weights of the energy and other terms affect how the model optimizes the energy cost, and the network provider can adjust the weights based on their demands. Although D3PG consumes more energy than DDPG-softmax, they consume almost the same amount of energy to process each task as shown in Fig.~\ref{fig:energy_to_task}, and the D3PG model can save more energy than DDPG and TD models. {\color{black}Interestingly, the greedy algorithm achieves the best results for tasks completion ratio, but it cannot stabilize the edge servers and plan the resource for long-term as shown in Fig.~\ref{fig:stable}.} 

% \begin{figure}
%   \centering
%   \includegraphics[width=3.5in]{figures/engery.pdf}
%   \caption{Energy Consumption}
%   \label{fig:energy}
% \end{figure}
% \begin{figure}
%   \centering
%   \includegraphics[width=3.5in]{figures/sum_engery.pdf}
%   \caption{Energy to Tasks}
%   \label{fig:energy_to_task}
% \end{figure}

The models also can reduce the time cost while maximizing the number of completed tasks and save energy, as shown in Fig.~\ref{fig:time}. Although the reward function has contained the number of completed tasks before expiring, reduce the time cost can improve the user experience. As we can see from the Fig.~\ref{fig:time}, the average time cost decrease as the models converge to the near-optimal policies. Fig.~\ref{fig:time} shows that the D3PG model saves more time than other models, which improves the quality of service. The DDPG and DDPG-softmax models do not learn to reduce the time cost because the weights of time consumption are relatively small. Note that both energy consumption and average time cost are for the total cost of the completed and expired tasks.
% \begin{figure}
%   \centering
%   \includegraphics[width=3.5in]{figures/mean_time.pdf}
%   \caption{Average Time Cost}
%   \label{fig:time}
% \end{figure}

Fig.~\ref{fig:stable} shows the stability of models, which is measured by the number of time steps of each epoch; the stability is measured by the number of steps that the MEC servers can persist in each episode. The task partitioning and offloading for the MEC can be considered continuous tasks from a reinforcement learning perspective; there is no endpoint of the offloading unless the servers are overloaded or crashed. However, for the simplicity of training, we formulate the task partitioning and offloading as episodic reinforcement learning task; therefore, {\color{black}we set the episodes to end when one of the edge servers is overloaded. In this simulation, we also add the external trigger to terminate the episodes, and the episode is forced to terminate when the number of time steps greater than a threshold value, set as 1,000.} Again, we run training the models for five times and average the results. As we can see from Fig.~\ref{fig:stable} TD3 and D3PG models can reach nearly 1,000 epochs, which indicates most of the epochs are stopped by the simulator, and they can maintain a stable MEC network. {\color{black}The greedy algorithm only chooses the action that maximizes the current reward for each time step but does not plan resources for the long term; therefore, the edge servers are easily overloaded under the greedy algorithm control.}

\section{Conclusions}
\label{sec_con}
In this work, we have studied task partitioning and computation offloading {\color{black}in a dynamic environment with multiple IoT devices and multiple edge servers. We developed an end-to-end DRL method to partition and offload tasks and allocate the edge servers' computing power to achieve joint optimization of expected long-term rewards. The model is optimized to maximize the completed tasks before the deadlines, minimizing energy consumption and simultaneously minimizing the time cost.  In order to deal with the constrained hybrid action space, we proposed a novel DRL model, namely D3GP, by integrating the Dirichlet distribution into DDGP to make decisions for task partitioning and the Ornstein-Uhlenbeck process for frequency control.} The developed model has been verified with extensive simulations and comparisons with existing methods, and the results show that our model outperforms state-of-the-art DRL models. For future work, we will study the optimization for task partitioning and offloading based on the sub-tasks that have dependency relations and task priorities.

\nocite{*}
\bibliographystyle{IEEEannot}
\bibliography{annot}
\end{document}